%% file: StabilityAndNonlocal-0.tex
\documentclass[11pt,english]{article}
\usepackage[english]{babel}
\usepackage{latexsym}
\usepackage[T1]{fontenc}
\usepackage{hyperref}
\usepackage[backend=bibtex,
            style=numeric-comp,
            sorting=none,
            sortcites=true,
            maxbibnames=99,
						doi=false
           ]{biblatex}

\usepackage{color}
\usepackage{amsmath,amsfonts,amssymb}
\usepackage{graphicx}
\DeclareGraphicsExtensions{.jpg,.pdf,.tif,.mps,.png}

\newcommand{\RR}{\ensuremath{ \mathbb{R}} }

\newcommand{\ZZ}{\ensuremath{ \mathbb{Z}} }

\newcommand{\D}{{\rm d}}

\newcommand{\realp}{{\mathbb{R}}{\rm e}}

\newtheorem{definition}{Definition}

\newtheorem{corollary}{Corollary}
\newtheorem{theorem}{Theorem}

\begin{document}
\title{Are nonlocal Lagrangian systems fatally unstable? }
\author{Carlos Heredia\thanks{e-mail address: carlosherediapimienta@gmail.com} \, and \, Josep Llosa\thanks{e-mail address: pitu.llosa@ub.edu}\\
Facultat de F\'{\i}sica (FQA and ICC) \\ Universitat de Barcelona, Diagonal 645, 08028 Barcelona, Catalonia, Spain }
\maketitle

\begin{abstract}
We prove that higher-derivative and genuinely nonlocal Lagrangian systems can be Lya\-punov-stable even when their Hamiltonians lack a lower bound. Explicit free and coupled Pais-Uhlenbeck oscillators, together with a genuine nonlocal model, are analysed to identify the precise conditions under which stability holds. These counterexamples point out the logical gap in the “Ostrogradsky instability” claims and provide benchmarks for constructing efficient stable higher-derivative theories.

\noindent
\end{abstract}

\section{Introduction \label{S1}}

Nonlocal and higher--derivative modifications of gravity have emerged as promising frameworks for explaining both the late--time acceleration of the Universe---without invoking an explicit cosmological constant---and key features of the inflationary epoch (see, e.g.,\ \cite{DeserWoodard2007,MaggioreMancarella2014,Biswas2012, Paul2017,Belgacem2018,Motohashi2020,Boumaza2020,Baptista2021,Joshi2022,Naruko2023}). These models have therefore attracted significant interest in theoretical cosmology. However, many researchers view such frameworks with suspicion, owing to concerns about a purported ``Ostrogradsky instability,” frequently described as a no--go ``theorem” that would doom any Lagrangian containing time derivatives beyond first order. 

The aim of this paper is to address what we believe to be a widespread misconception. We argue that what is often labeled the “Ostrogradsky instability” (or also known as “Ostrogradsky theorem”) is a misunderstanding that would not, in itself, warrant a dedicated article. Yet its repeated appearance in the literature compels us to address it explicitly. Although several authors have already sought to clarify the matter—for instance, Smilga’s study of higher‑derivative theories and ``benign ghosts”\cite{Smilga2017}, the discussion by Defayet et al.\ \cite{Defayet}, or the recent treatment by Errasti et al.\ \cite{ErrastiDiez2025}—confusion persists.

A search for ``Ostrogradsky theorem'' typically yields two primary results. The first is the well-known Divergence Theorem, also known as the Gauss-Ostrogradsky Theorem, whose significance is unquestionable. The second result, which is the main focus of our extensive commentary here, relates to the Ostrogradsky M\'emoire \cite{Ostrogradsky}.
According to one interpretation \cite{OstroWiki, Sarrion2020, Aoki2020}, it states:
\begin{quote}
``A non-degenerate Lagrangian dependent on higher-than-first-order time derivatives corresponds to a Hamiltonian that is unbounded from below.''
\end{quote}
In contrast, another interpretation \cite{Woodard1989,Woodard2015,Woodard2007, Fasiello2013} suggests:
\begin{quote}
``In any non-degenerate theory with a fundamental dynamical variable of higher than second-order time derivative, there exists a linear instability.''
\end{quote}
The first interpretation indeed represents a theorem, albeit a relatively straightforward one. The latter, however, is incorrect and thus cannot be considered a theorem, as we will demonstrate through various counterexamples.
In this context, several authors \cite{OstroWiki, Aoki2020, Woodard2015, Woodard2007, Fasiello2013,  GanzNoui2021, Donohue2021} refer to the ``Ostrogradsky instability.'' Notably, there is no documentary evidence directly linking Ostrogradsky to these assertions, although his M\'emoire \cite{Ostrogradsky} is frequently cited. A thorough examination of this source reveals no such statements. In fact, obtaining the document is not an easy task, and it is cumbersome to read for a contemporary scholar because of the archaic notation, more closely related to that of Lagrange and Poisson \cite{Lagrange1788, Poisson} than to the notation of Jacobi and Hamilton \cite{Jacobi, Hamilton}, which has ended up being imposed \cite{Fraser2023}.
Ostrogradsky does not refer to the ``Hamiltonian'' ---he does not mention the prior work of Hamilton and Jacobi, though he declares that he knows of the existence of a letter from the latter, which he says he has not read. Moreover, Ostrogradsky's work does not address the function of the Hamiltonian in this context, nor the notion of being unbounded from below. 

In our view, inferring that a system is unstable from the fact that the Hamiltonian is not bounded from below is nothing but an instance of the so-called {\em fallacy of the converse} or {\em affirming the consequent} \cite{contrareciproc}. While the presence of a strict energy minimum is a sufficient condition for stability (as per the Lagrange-Dirichlet theorem \cite{Lagrange1788}), it is not a necessary condition. We will elaborate on this in Section \ref{S2},  where the definition of Liapunov stability, the theorem of Dirichlet-Lagrange and the direct method of Liapunov, as sufficient conditions for stability, are outlined. 
 
In Section \ref{S3}, these notions are applied to the second-order Pais-Uhlenbeck (PU) oscillator \cite{Pais1951}, which is the system considered by most authors addressing the issue of stability in higher-order and nonlocal Lagrangian systems, see for instance \cite{Aoki2020,GanzNoui2021}. We will consider it here either free or in interaction, either internally or externally. We do not consider specifically higher-(finite-)order PU oscillators because the treatment is quite similar. 

In Section \ref{S4}, we study the stability of a genuinely nonlocal system, an infinite-order Pais-Uhlenbeck system. In almost all examples we study, Liapunov's Corollary is used as a sufficient condition indicating stability, in spite of the fact that the Hamiltonian is not positive definite.


\section{The theory of stability: a brief outline  \label{S2}}
Here we outline the main notions of stability of an equilibrium position as presented in ref. \cite{Gantmacher} from Liapunov's work \cite{Liapunov}.

\begin{definition}[Liapunov stability] \label{D1} 
Let $q_j = \dot q_i = 0 \,,\;\, i,j = 1 \ldots n$ be an equilibrium position
of a dynamical system. It is said to be stable if for ``small enough'' initial deviations, $\,q_{j0}\,,\;\dot q_{i0}\,$, the system does not leave a ``small'' neighborhood of the equilibrium position, that is $\forall\; \varepsilon > 0\,$, it exists $\delta(\varepsilon) > 0\,$  such that 
$$ \sup\{|q_{j0} |,|\dot q_{i0}|\}_{i,j=1...n} < \delta
\qquad \Rightarrow \qquad  \sup \{|q_{j}(t) |,|\dot q_{i}(t)|\}_{i,j=1...n} < \varepsilon \,, \quad \forall\;t > 0 \,, $$
\end{definition}
where \textit{sup} denotes the supremum. The equilibrium position is said to be {\em asymptotically stable} if the ``small enough'' initial deviation tends to vanish for $\,t \rightarrow \infty\,$, that is
$$ \sup \{|q_{j0} |,|\dot q_{i0}|\}_{i,j=1...n} < \delta
\qquad \Rightarrow \qquad  \lim_{t\rightarrow\infty} q_{j}(t)= 0 \,. $$

There is a variety of theorems concerning the stability of an equilibrium position. The most
popular among them is \cite{Lagrange1788}:
\begin{theorem}[Lagrange] \label{T1}
If the potential energy of a conservative system has a strict minimum at a given position, then this is a stable equilibrium point\footnote{It was rigorously proved by Dirichlet \cite{Dirichlet}.}.
\end{theorem}
Since kinetic energy is generally non-negative, an immediate  Corollary follows from replacing the potential energy with the energy, $E\,$. 

Later Liapunov \cite{Liapunov,Liapunov0} generalized Lagrange theorem by replacing $\,E\,$ with a function $\,V (q,\dot q)\,$ that: (a) is continuously derivable, (b) has a strict minimum at the equilibrium position and (c) is not increasing under any motion, namely
\begin{theorem}[Liapunov] \label{T2}
 Let $q_j = \dot q_i = 0 \,,\quad i,j = 1 \ldots n$ be an equilibrium position of an autonomous system and let $\,V (q,\dot q)\,$ 
be a class $\mathcal{C}^1$ function which has a strict minimum at $\,q_{j0}=\dot q_{i0}=0 \,$,
whereas its total time derivative,  $\,\dot V (q,\dot q, \ddot q)\,$, has a maximum.
Then the equilibrium is stable. Furthermore, if the maximum of $\dot V$ is an strict maximum, then the equilibrium position is asymptotically stable.
\end{theorem}
The function $\,V (q,\dot q)\,$ is usually referred to as {\em Liapunov function}.
A particular instance of Theorem \ref{T2} is the following: 
\begin{corollary}[Liapunov]  \label{C1}
If $\,V (q,\dot q)\,$  is an integral of motion and it has a strict minimum at $q_j = \dot q_i = 0 \,,\,\, i,j = 1 \ldots n$, then it is a stable equilibrium point.
\end{corollary}
Both theorems and the corollary establish conditions that are sufficient, but not necessary, for stability. Deriving general necessary conditions is considerably more challenging, except in the case of linear systems \cite{Liapunov1}—and even more so when those systems have constant coefficients, as
\begin{equation} \label{SLC1}
\dot{\mathbf{x}} = \mathbb{M}\,\mathbf{x} \,, \qquad {\rm where} \qquad 
\mathbf{x} = \left(\begin{array}{c}
 x_1 \\ x_2 \\ \vdots 
\end{array} \right)  \,,
\end{equation}
$\,\mathbb{M}\,$ is a constant square $n\times n$ matrix and the variables $x_j$ stand for the coordinates $q$ and velocities $\dot q\,$ of the system. It has one equilibrium point, namely $\,x_k = 0\,,\,\, k = 1\ldots n\,$ .
\begin{theorem} \label{T3}
{\em (a)} If the real parts of all characteristic roots of $\,\mathbb{M}$ are negative, then the equilibrium point is asymptotically stable and {\em (b)} if $\,\mathbb{M}$ has a characteristic root whose real part is positive, then it is unstable \cite{Liapunov1}.
\end{theorem}

Next theorem considers the case that some among the characteristic roots are imaginary.
\begin{theorem} \label{T4}
{\em (a)} If the characteristic roots of $\mathbb{M}$ either have negative real part or are imaginary and simple, then the equilibrium point is stable and {\em (b)} 
if either the real part of one among the characteristic roots is positive or one of the characteristic roots is imaginary and multiple, then it is unstable.
\end{theorem}
See Appendix A for the details of the proof.

\section{The Pais-Uhlenbeck oscillator  \label{S3}}
We study the second-order Pais-Uhlenbeck oscillator under three different scenarios. First, we examine the free case; next, we consider its interaction with an external oscillator; and finally, we analyze the case where an internal interaction occurs between the oscillator’s two degrees of freedom.
Our aim is to show that a large class of values of the coupling constant lead to a stable behavior, despite the fact that the Lagrangian contains second-order derivatives.

\subsection{The free Pais-Uhlenbeck oscillator.  \label{FPU}}
The most cited example in the literature is the second-order PU oscillator \cite{Pais1951}, which is ruled by the fourth order Lagrangian 
\begin{equation}  \label{FPU0}
-\frac12\,q\,\left(1+\frac{D^2}{\omega_1^2}\right)\,\left(1+\frac{D^2}{\omega_2^2}\right)\,q\,, 
\end{equation}
or, equivalently, by the second-order Lagrangian
\begin{equation}  \label{FPU1}
L_0 = -\frac12 \,q^2 + \frac12\,\left(\frac1{\omega_1^2}+\frac1{\omega_2^2}\right) \dot q^2 - \frac1{2 \omega_1^2 \omega_2^2}\,\ddot q^2
\,, \qquad \omega_1^2 < \omega_2^2  \,,
\end{equation} 
apart from boundary terms. To set up the Hamiltonian formalism, we implement the Ostrogradsky procedure and write $\; X_1 := q\,, \quad X_2 := \dot q  \,,$
whereas the respective conjugate momenta are given by the Legendre-Ostrogradsky transformation
\begin{equation}  \label{FPU2a}
P_2 := \frac{\partial L}{\partial \ddot q} = - \frac1{\omega_1^2 \omega_2^2} \ddot q \,, \qquad 
P_1 := \frac{\partial L}{\partial \dot q} - \frac{\D P_2}{\D t} = \frac{(\omega_1^2 + \omega_2^2)\,\dot q + \stackrel{\ldots}{q}}{\omega_1^2 \omega_2^2}   \,,
\end{equation}
whose inverse  is
\begin{equation}  \label{FPU2b}
\ddot q = - \omega_1^2 \omega_2^2 \,P_2\,, \qquad \stackrel{\ldots}{q} = \omega_1^2 \omega_2^2\,P_1 - (\omega_1^2 + \omega_2^2)\,X_2 \,.
\end{equation}
The non-vanishing elementary Poisson brackets and the Hamiltonian are respectively
\begin{equation}  \label{FPU3}
\{X_a,P_b\} = \delta_{ab}  \qquad {\rm and} \qquad 
H = P_1 X_2 - \frac{\omega_1^2 \omega_2^2}2\,P_2^2 - \frac{\omega_1^2 +\omega_2^2}{2 \omega_1^2 \omega_2^2}\,X_2^2 + \frac12\,X_1^2   \,.
\end{equation}
Now, following ref. \cite{Pais1951} we introduce the new coordinates and momenta 
\begin{eqnarray}  \label{FPU4}
 q_1 = \frac{\omega_2\,\left(X_1-\omega_1^2 P_2\right)}{\omega_1\,\sqrt{\omega_2^2 - \omega_1^2}}\,,\hspace*{.5em} &\hspace*{3em}&  
q_2 = \frac{\omega_1\,\left(X_1-\omega_2^2 P_2\right)}{\omega_2\,\sqrt{\omega_2^2 - \omega_1^2}} \,,  \\[1.5ex]  \label{FPU4a}
p_1 = - \frac{\omega_1\,\left(X_2-\omega_2^2 P_1\right)}{\omega_2\,\sqrt{\omega_2^2 - \omega_1^2}}  \,, &\hspace*{3em}& 
p_2 = \frac{\omega_2\,\left(X_2-\omega_1^2 P_1\right)}{\omega_1\,\sqrt{\omega_2^2 - \omega_1^2}} \,.
\end{eqnarray}  
They are also a canonical set, $\; \{q_a,q_b\} =  0  \,, \;\, \{q_a,p_b\} =  \delta_{ab}  \,, \;\,  \{p_a,p_b\} =  0  \,$, and
in terms of them the Hamiltonian becomes 
\begin{equation}  \label{FPU5}
H = \frac12\,\left(p_1^2 + \omega_1^2 q_1^2\right) - \frac12\,\left(p_2^2 + \omega_2^2 q_2^2\right)  \,.
\end{equation}

\subsubsection*{Stability}
The Lagrange theorem does not allow to infer the stability of the equilibrium position $\,q_a = p_b =0\,$ because the Hamiltonian has no definite sign. However, Corollary \ref{C1} in Section \ref{S2} can be invoked to prove that it is stable.
Indeed, the Hamilton equations are
$$  \dot q_1 = p_1  \,, \qquad \dot q_2= - p_2 \,,\qquad     
\dot p_1 = -\omega_1^2\,q_1 \,,\qquad\dot p_2 = \omega_2^2\,q_2  \,, $$
whence it follows that
\begin{equation}  \label{FPU6}
  \ddot q_a + \omega_a^2\,q_a = 0 \,, \qquad  a =1,2\,, 
\end{equation}
that is two harmonic oscillators with frequencies $\omega_a\,$. Particularly, the individual energies of both oscillators,
$\, h_a := \frac12\,\left(p_a^2 + \omega_a^2 q_a^2\right) \,, \quad a= 1, 2 \,$, are integrals of motion and so it is their sum $\,G:= h_1 + h_2\,$, which is positive definite and has a strict minimum at the equilibrium point $\,q_a = p_b =0\,$. Hence the stability of the equilibrium point follows from Corollary \ref{C1}. 

A way of explaining this stability in spite of the unboundedness of energy consists of realizing that, as the subsystems $(q_1,p_1)$ and $(q_2,p_2)$ are decoupled, the energy of the first cannot become more positive at the cost of making that of the second more negative, because energy cannot be transferred from $(q_2,p_2)$ to $(q_1,p_1)\,$. Some authors argue that if the system is perturbed by adding some kind of interaction between both subsystems, then this energy transfer will be possible, which will cause instability. 

The aim of the following examples is to show that for a non-negligible range of interactions, either external (Section 3.2) or internal (Section 3.3), the system remains stable.

\subsection{The Pais-Uhlenbeck oscillator with an external interaction.  \label{EIPU}}
Now consider a PU oscillator (\ref{FPU1}) interacting with an external harmonic oscillator 
\[\, L_1 = \frac12\, m \,\left(\dot x^2 -\omega_0^2 x^2\right) \,\]
through an interaction term $\; L_{\rm int} = g \,\dot q\, x\,$. This is a toy model for a ``monochromatic  electromagnetic field'' $x$ interacting with a charged PU oscillator. The total Lagrangian is 
\begin{equation}  \label{PU2}
L = -\frac12 \,q^2 + \frac12\,\left(\frac1{\omega_1^2}+\frac1{\omega_2^2}\right) \dot q^2 - \frac1{2 \omega_1^2 \omega_2^2}\,\ddot q^2 + \frac12\, m \,\left(\dot x^2 -\omega^2 x^2\right) + g \,\dot q\, x \,,
\end{equation} 
which is second-order with respect to $q$ and first order with respect to $x$. To implement the Ostrogradsky procedure, we write $\; X_1 := q\,, \;\, X_2 := \dot q  \;\, {\rm and} \;\, X_0 := x \,$,
and the respective conjugate momenta are
\begin{eqnarray}  \label{PU2a}
P_2 &:=&  - \frac1{\omega_1^2 \omega_2^2} \ddot q \,, \hspace*{9em} \ddot q = - \omega_1^2 \omega_2^2 \,P_2\,, \\[1.5ex] \label{PU2b}
P_1 &:=&  \frac{\omega_1^2 + \omega_2^2}{\omega_1^2 \omega_2^2}\,\dot q + \frac1{\omega_1^2 \omega_2^2} \,\stackrel{\ldots}{q} + g x  \,, \quad\; \stackrel{\ldots}{q} = \omega_1^2 \omega_2^2\,P_1 - (\omega_1^2 + \omega_2^2)\,X_2 - g X_0\,,
\\[1.5ex]\label{PU2c}
P_0 &:=&  m\, \dot x \,, \hspace*{11em}\dot x = \frac{P_0}{m}\,.
\end{eqnarray}
The non-vanishing Poisson brackets are $\;\{X_a,P_a\} =  1  \,, \;\, a= 0, 1, 2 \,,$
and the Hamiltonian is
\begin{equation}  \label{PU3}
H = P_1 X_2 - \frac{\omega_1^2 \omega_2^2}2\,P_2^2 + \frac{P_0^2}{2 m} - \frac{\omega_1^2 +\omega_2^2}{2 \omega_1^2 \omega_2^2}\,X_2^2 + \frac12\,X_1^2 + \frac{m \omega_0^2}2\,X_0^2 - g\,X_2\,X_0   \,.
\end{equation}
Similarly as in Section \ref{FPU} we introduce the Pais-Uhlenbeck coordinates and momenta
\begin{equation}  \label{PU4}
q_0 = X_0\,, \qquad q_1 = \frac{\omega_2\,\left(X_1-\omega_1^2 P_2\right)}{\omega_1\,\sqrt{\omega_2^2 - \omega_1^2}}\,,\qquad  
q_2 = \frac{\omega_1\,\left(X_1-\omega_2^2 P_2\right)}{\omega_2\,\sqrt{\omega_2^2 - \omega_1^2}} \,,  
\end{equation}
\begin{equation}  \label{PU4a}
p_0 = P_0 \,, \qquad  p_1 = - \frac{\omega_1\,\left(X_2-\omega_2^2 P_1\right)}{\omega_2\,\sqrt{\omega_2^2 - \omega_1^2}}  \,, \qquad
p_2 = \frac{\omega_2\,\left(X_2-\omega_1^2 P_1\right)}{\omega_1\,\sqrt{\omega_2^2 - \omega_1^2}}  \,,
\end{equation}
whose Poisson brackets vanish, except $\,\{q_a,p_a\} = 1\,, \quad a= 0, 1, 2 \,$.

In terms of these coordinates the Hamiltonian reads
\begin{equation}  \label{PU5}
H = \frac12\,\left(p_1^2 + \omega_1^2 q_1^2\right) - \frac12\,\left(p_2^2 + \omega_2^2 q_2^2\right) + \frac{p_0^2}{2 m} + \frac{m \omega_0^2}2\,q_0^2 + H_{\rm int} \,,
\end{equation}
where
$$ H_{\rm int} = - \hat{g} q_0\,(p_1+p_2) \,, \qquad {\rm with} \qquad  
\hat{g}:=\frac{g \,\omega_1 \omega_2}{\sqrt{\omega_2^2-\omega_1^2}}  \,. $$

Although the Hamiltonian has not a definite sign, we will show that the system is stable for a wide range of values of the coupling constant $g$. Indeed, the Hamilton equations are:
\begin{eqnarray}  \label{PU6a}
\dot q_1 = p_1 - \hat g q_0 \qquad & \qquad \dot q_2 = -p_2 - \hat g q_0 \qquad  & \qquad \dot q_0 = \frac{p_0}m \\[1.5ex]  \label{PU6b}
\dot p_1 = - \omega_1^2\,q_1 \qquad & \qquad \dot p_2 = \omega_2^2\,q_2 \qquad & \qquad \dot p_0 = - m \omega_0^2 q_0 + \hat g\,(p_1+p_2) \,.
\end{eqnarray}
This is linear differential system that can be written in matrix form as
\begin{equation}  \label{PU6}
\frac{\D \mathbf{x}}{\D t} = \mathbb{M}\,\mathbf{x} \,,
\end{equation}
where the matrix and vector respectively are
\begin{equation}  \label{PU7}
\mathbb{M} = \left(\begin{array}{ccc|ccc}
                   0 & 0 & -\hat g & 1 & 0 & 0 \\
                   0 & 0 & - \hat g & 0 & -1 & 0 \\
                   0 & 0 & 0      & 0 & 0 & 1/m \\
									\hline
              -\omega_1^2 & 0 & 0   & 0 & 0 & 0 \\
              0 & \omega_2^2 & 0   & 0 & 0 & 0 \\
              0 & 0 & -m\omega_0^2 & \hat g & \hat g & 0 
							     \end{array} \right)		
 \qquad {\rm and} \qquad
\mathbf{x} = \left(\begin{array}{c}
                   q_1\\ q_2 \\ q_0 \\ p_1 \\ p_2 \\ p_0
							     \end{array} \right)  \,.	
\end{equation}
The characteristic determinant is
$$ \chi_\mathbb{M}(\lambda) := \det (\mathbb{M}- \lambda \mathbb{I}) \equiv (\omega_0^2 +\lambda^2)(\omega_1^2 +\lambda^2)(\omega_2^2 +\lambda^2) + \gamma \,\lambda^2 \,, $$
with 
$$\,\gamma:= \frac{\hat g^2 (\omega_2^2 - \omega_1^2) }m =  \frac{g^2\,\omega_1^2 \omega_2^2 }m  > 0  \,. $$

As seen in Theorems \ref{T3} and \ref{T4} in Section \ref{S2}, the stability of the system is indicated by the kind of characteristic roots. Since $ \chi_\mathbb{M}(\lambda)$ is an even function of $\lambda$, the roots come in pairs, $\pm\lambda_j$, and the condition $\,\realp(\pm\lambda_j) \leq 0\,, \;\,\forall \lambda_j\,$, amounts to $\,\realp (\pm\lambda_j) = 0\,$ or, equivalently, 
$ \, \lambda_j^2 < 0\,$. (The case $ \, \lambda_j = 0\,$ is excluded because it would imply that 0 is a multiple root of $ \chi_\mathbb{M}(\lambda)\,$, which carries instability.)

This implies that, if the polynomial $\,P(Q):= (\omega_0^2 +Q)(\omega_1^2 +Q)(\omega_2^2 +Q) + \gamma \,Q\;$ has three simple negative roots, then the system (\ref{PU6}) is stable.

Now, since $\gamma >0\,$, $\,P(Q)> 0\,$ for all real non-negative $\,Q\,$, therefore any real root must be negative and it is enough to require that all roots of the cubic polynomial $\,P(Q)\,$ are real and simple or, equivalently, that the discriminant is negative. 

The polynomial can be written as $\,P(Q) \equiv Q^3 + a_1 Q^2 + (\gamma +a_2) Q + a_3\,$, where the $a_l$'s depend on the $\omega$'s, 
$$ a_1 = \omega^2_1+\omega^2_2+\omega^2_0  \,, \qquad a_2 = \omega^2_0 \left(\omega^2_1+\omega^2_2 \right) +\omega^2_1+\omega^2_2 \qquad {\rm and} \qquad  a_3 = \omega^2_1 \omega^2_2 \omega^2_0 \,, $$
and the discriminant is
$$ D(\gamma) = \frac1{27}\,\gamma^3 + b_1 \gamma^2 + b_2 \gamma + b_3\,,$$  
where the coefficients $b_1,b_2$ and $b_3$ depend on the $\omega$'s only.
Notice that $\displaystyle{ \lim_{\gamma\rightarrow \infty} D(\gamma) \rightarrow +\infty}\,$ and that $D(0) < 0\,$ because, for $\gamma =0$, the polynomial $P(Q)$ has three simple negative roots, namely $-\omega_0^2\neq -\omega_1^2 \neq -\omega_2^2\,$. Therefore at least a positive $ \gamma_1\,$ exists such that $\,D(\gamma_1) =0\,$ and $\,D(\gamma) < 0\,$ for all $\,0\leq \gamma < \gamma_1\,$. Thus for any coupling constant $\,\displaystyle{|g| < \frac{\sqrt{m \gamma_1}}{\omega_1\omega_2} }\,$, the polynomial $P(Q)$ has three simple negative roots and  the system is stable. 

\subsection{The Pais-Uhlenbeck oscillator with an internal interaction}
Are there internal interactions connecting both oscillators of a PU system such that stability is preserved?
Let  
\begin{equation}\label{PUI0}     
 H = h_1-h_2 + V(q,p) \,, \qquad h_a =\frac12\,\left(p_a^2 + \omega_a^2 q_a^2 \right) \,
\end{equation}
be a perturbation of the PU Hamiltonian (\ref{FPU5}). We look for a Liapunov function $G(q,p)\,$, that is (a) an integral of motion $\; \{H,G\} =0 \,$ and (b) it has a strict minimum at $q_a=p_b=0\,$.

Writing the coordinates as $\,\left(x_i \right)_{i=1\ldots4} = (q_1,q_2,p_1,p_2)\,$, the condition that $G$ is an integral of motion implies that
\begin{equation}\label{PUI1}     
\partial_i H \,\Omega_{ij} \,\partial_j G = 0\,, 
\end{equation}
where summation over repeated indices is understood and all $\Omega_{ij}$ vanish except
$$ \Omega_{13} = \Omega_{23} = -\Omega_{31} = -\Omega_{42} = 1  \,.$$
In the case that $V$  is a homogeneous quadratic function, namely $\displaystyle{ V = \frac12\,V_{ij} x_i x_j}\,$, the Hamiltonian is quadratic too and equation (\ref{PUI1}) admits a particular solution that is also quadratic $\displaystyle{ G = \frac12\,G_{ij} x_i x_j}\,$. Obviously, $\,V_{ij} = V_{ji}\,$, $\,G_{ij} = G_{ji}\,$ and $\,H_{ij} = H_{ji}\,$, and  the equation reads
$$ H_{il}\,\Omega_{ij} \,G_{jk} \,x_l x_k	= 0\,, \qquad \forall x_l \,,$$		
which amounts to						
\begin{equation}\label{PUI2}     
H_{li}\,\Omega_{ij} \,G_{jk} + H_{ki}\,\Omega_{ij} \,G_{jl} = 0 \,. 
\end{equation}									
Moreover, in the present quadratic case, the  strict minimum condition is equivalent to require the positivity of $\,G(x)\,$ for $x_j\neq 0\,$.

A Liapunov function $G$, solution of (\ref{PUI2}), exists for many interaction potentials. In Appendix B we show that a solution exists for the quadratic potential $\,\displaystyle{V(q) = \frac{a}2 q_1^2 + \frac{b}2 q_2^2 + c\,q_1 q_2 }\,$, as long as the coefficients fulfill the inequalities 
\begin{equation}  \label{PUI90} 
\left(a + b + \omega_1^2 -\omega_2^2 \right)^2 > 4 c^2 >  4 \,\left(a + \omega_1^2\right)\,\left(b -\omega_2^2 \right) \qquad {\rm and } \qquad   a + \omega_1^2  > b -\omega_2^2  \,
\end{equation}
and the Liapunov function is 
\begin{equation}\label{PUI91} 
 G = h_1 + h_2 + V - \frac{c^2 q_1^2 - \left[c^2 + b(a + b +\omega_1^2 -\omega_2^2)\right]\,q_2^2 + 2 c \left[b-\omega_2^2 \right] q_1 q_2 + 2 c p_1 p_2}{a + b +\omega_1^2 -\omega_2^2}  \,. 
\end{equation}
As shown in Appendix B, the integral of motion $G$ has a strict minimum at $p_a = q_b=0\,$, therefore Liapunov's Corollary \ref{C1} ensures the stability of the system with the interaction potential $V(q)$ whenever the inequalities (\ref{PUI90}) are satisfied. 

\input{GhostsNoRunaways.tex}


\section{Genuine nonlocality. } \label{S4}
So far we have considered  finite order Pais-Uhlenbeck oscillators, either free or interacting ---actually only second-order Lagrangians---, but for higher orders the technique is the same and the results are very similar. We will now study a system that is genuinely nonlocal, namely a PU oscillator of infinite order.

In the local case, Lagrange equations form an ordinary differential system of order $2 n\,$, whereas in the nonlocal case the equations include terms that depend on entire segments of the trajectory. As a consequence, while in the local case a finite number of initial data, i. e. the coordinates and their derivatives up to order $2 n-1$, are enough to determine the future evolution, for nonlocal equations it is not so and, as a rule, an infinite number of data $\,(\alpha_j)_{ j \in \mathcal{J}}\,$ are needed to determine a solution and often these data do not belong to a fixed instant of time. With this in mind, Definition \ref{D1} for stability must be adapted to nonlocal equations of motion.

\begin{definition}  \label{D3}
Given a nonlocal system having an equilibrium position $\,q_0\,$ that corresponds to the data $\,(\alpha_{0j})_{ j \in \mathcal{J}}\,$, that is  $q(t,\alpha_{0j}) = q_0\,$, this equilibrium position is said stable if (for some given norm in the space of data)
$\forall \varepsilon >0\,$, it exists $\delta(\varepsilon) >0$ such that $\|\alpha_j - \alpha_{0j} \| < \delta\,$ implies that  $\left|q(t,\alpha_{j}) - q_0)\right| < \delta\,$, $\forall t\,$.
\end{definition}

Next we study the example of a fully nonlocal Pais-Uhlenbeck oscillator. We will see that the ``initial'' data are a bounded piece of the whole trajectory, but it could be more complex in other cases, e. g. the $p$-adic string \cite{pstring}. We emphasize that this example is provided purely for illustrative purposes, and is not intended to represent a physical system (even if one might exist).
 
When the operator $F(D)$ in the Lagrangian (\ref{FPU0}) is an infinite product, 
\begin{equation}  \label{PUNL0} 
L = - q\,F(D) q\,, \qquad \quad F(D) =\prod_{n=1}^\infty \left(1 + \frac{D^2}{\omega_n^2}\right)\,,
\end{equation} 
then it is completely nonlocal. For the reasons given in \cite{IDvsOI}, instead of dealing with an infinite number of derivatives, we here adopt the functional formalism for nonlocal systems and convert the operator $F(D)$ into an integral operator through the Fourier transform 
\begin{eqnarray*}
F(D) q(t) &:=& \frac1{\sqrt{2 \pi}}\,F(D) \int_\RR \D k\,e^{-i k t}\,\tilde{q}(k) = \frac1{\sqrt{2 \pi}}\,\int_\RR \D k\,F(D) e^{-i k t}\,\tilde{q}(k) \\[1ex]
 & = & \frac1{\sqrt{2 \pi}}\,\int_\RR \D k\,\prod_{n=1}^\infty \left(1 + \frac{(- i k)^2}{\omega_n^2}\right)\,\, e^{-i k t}\,\tilde{q}(k)\,.
\end{eqnarray*}
By the convolution theorem we then have that
\begin{equation}  \label{PUNL00}
F(D) q = K \ast q \,, \qquad {\rm with}  \qquad K(t) := \frac1{2 \pi}\,\int_\RR \D k\,  F(- i k) \, e^{-i k t}  \,,
\end{equation}
where the asterisc means the convolution, and the function in the integral is the infinite product 
\begin{equation}  \label{PUNL00a}
 F(-ik):=\prod_{n=1}^\infty \left(1 - \frac{k^2}{\omega_n^2}\right)\,. 
\end{equation}
A necessary and sufficient condition for the infinite product to be absolutely convergent is that the series $\,\displaystyle{ \sum_{n=1}^\infty \frac{1}{\omega_n^2} }\,$ is convergent ---see \cite{Whittaker}. Hence, the infinite product (\ref{PUNL00a}) converges uniformly for any complex value of $k\,$ and $F$ is an entire function.

Here we will consider in detail the particular case $\,\omega_n = \pi\,n\,$ and use the infinite product \cite{Whittaker}
$$ F(z) = \prod_{n=1}^\infty \left(1 + \frac{z^2}{\pi^2 n^2}\right) = \frac{\sinh z}{z} = \sum_{n=0}^\infty \frac{z^{2n}}{(2n+1)!} \,.$$
Applying the definition (\ref{PUNL00}), we have that
\begin{equation}  \label{PUNL1a}  
K(t) := \frac1{2 \pi}\,\int_\RR \D k\, \frac{\sin k}{k} \, e^{-i k t} = \frac12\,\Theta(1-|t|) \,, 
\end{equation}
therefore $F(D)$ is the convolution operator 
\begin{equation}  \label{PUNL1}
F(D) q = K \ast q \,, \qquad {\rm with}  \qquad K(t) = \frac12\,\Theta(1-|t|)  \,
\end{equation}
and the action integral for the Lagrangian (\ref{PUNL0}) is ---see ref. \cite{Heredia2022a} for details---
$$ S(q) = -\int_{\RR} \D t\, \int_{\RR} \D\tau\,q(t)\,q(\tau)\,K(t-\tau) \,.$$
The variational principle then reads $ \;\delta S(q) = 0\,$, for any variation $\,\delta q\,$ of compact support, and it leads to the Lagrange equation $\,  K \ast q = 0\,$, which can be written as
\begin{equation}  \label{PUNL2}
 K \ast q = 0 \qquad {\rm or}  \qquad \int_{-1}^1 \D\tau\,q(t+\tau) = 0 \,, \qquad \forall t \in \RR \,,
\end{equation}
where (\ref{PUNL1a}) has been included.

It is obvious that $q(t)=0\,$ is a solution and $0$ is an equilibrium point. In order to assess its stability on the basis of Definition \ref{D3}, we need to identify a convenient space of data and choose a norm on it.
We can expect that $q(t)$ is a real continuous function, at least; therefore the integral is differentiable and equation (\ref{PUNL2}) is equivalent to
\begin{equation}  \label{PUNL2b}
q(t+1) = q(t-1) \qquad {\rm and}  \qquad \int_{-1}^1 \D\tau\,q(\tau) = 0 \,,
\end{equation}
i. e. the solutions of (\ref{PUNL2}) are continuous periodic functions, with period 2 and vanishing mean value. Hence each solution is determined by the ``initial data'' $\,Q \in \mathcal{C}^0([-1,1])\,$ such that
$$ Q(-1)=Q(1)  \qquad {\rm and }\qquad \int_{-1}^1 \D\tau\,Q(\tau) = 0 \,,$$
which evolve into the future (and backwards into the past) as 
\begin{equation}  \label{PUNL2c}
\, q(\tau + 2 n) = Q(\tau) \,, \qquad n \in \ZZ \qquad {\rm and }\qquad |\tau| \leq 1 \,. 
\end{equation}

Since our analysis of stability will involve the energy, which depends on the velocity, we will also require $q(t)$ to be continuously differentiable and take the data space as
$$ \mathcal{D}:= \left\{Q \in \mathcal{C}^1([-1,1])\,|\;Q(-1)=Q(1)\,, \;\;\mbox{with vanishing mean value} \right\} \,, $$ 
and the norm
$\,\displaystyle{ \| Q\|_\mathcal{D} = \sup_{|\tau|\leq 1} \left| Q(\tau) \right|} \,$.

Any small displacement $\delta q $ from the equilibrium $ q=0\,$ will remain small because, according to the evolution law (\ref{PUNL2c}),
$$ \left|\delta q(\tau + 2 n) \right| \leq \sup_{|\tau|\leq 1} \left|\delta  Q(\tau) \right| = \| \delta Q\|_\mathcal{D} \, $$
and therefore the equilibrium is stable. (Definition \ref{D3} with $\delta(\varepsilon) = \varepsilon\,$).

We will now see that the energy has no definite sign despite the stability. 
Given the Lagrangian (\ref{PUNL0}), the energy for the solution $q $ corresponding to the data $Q \in \mathcal{D}$ is ---see \cite{Heredia2022a}, eq. (43)---
\begin{equation}   \label{PUNL3a} 
h(q) = \int_{\RR^2} \D\rho\,\D\sigma \,\left[\Theta(\sigma) - \Theta(\rho) \right]\, \dot q(\sigma) \,
\frac{\delta L(T_\tau q)}{\delta q(\sigma)} - L(q) \,,  
\end{equation}	
where $T_\tau q (\rho) := q(\tau+\rho) \,$ and 
\begin{equation}   \label{PUNL3} 
 \frac{\delta L(T_\tau q)}{\delta q(\sigma)} = - \delta(\tau - \sigma)\,K\ast q(\tau) -q(\tau)\,K(\sigma-\tau)  \,.  
\end{equation}	
The term $L(q)$ does not contribute because $q$ is a solution of the Lagrange equations and $K\ast q = 0\,$. To calculate the integral in the first term, we substitute (\ref{PUNL3}) into (\ref{PUNL3a}) and obtain
$$  h(q) = - \int_{\RR^2} \D\tau\,\D\sigma \,\left[\Theta(\sigma) - \Theta(\tau) \right]\, \dot q(\sigma) \,q(\tau) \,K(\sigma-\tau)\,$$
and, including that 
$$ \int_{\RR^2} \D\tau\,\D\sigma\,f(\sigma,\tau)\,\left[\Theta(\sigma) -\Theta(\tau)\right] =  \int_{\RR^{+2}} \D\tau\,\D\sigma\,
\left[f(\sigma,-\tau) -f(-\sigma,\tau)\right]  \,, $$
it yields
$$ h(q) = - \frac12\, \int_{\RR^{+2}} \D\tau\,\D\sigma \,\left[ \dot q(\sigma) \,q(-\tau) - \dot q(-\sigma) \,q(\tau) \right]\,
\Theta\left(1 -\sigma -\tau \right)\,,$$
where (\ref{PUNL1}) has been included. 
Notice that the Heaviside function cuts off any contribution of the part of the integrand with $|\sigma|>1 $ or $|\tau|>1 $, therefore the energy $h(q)$ only depends on the initial data $Q\,$ and
\begin{eqnarray}
  h &=& - \frac12\,\int_0^1 \D\tau\,\int_0^{1-\tau} \D\sigma \,\left[\dot Q(\sigma) \,Q(-\tau) - \dot Q(-\sigma) \,Q(\tau) \right]\nonumber \\[1.5ex]
	 &=& \frac12\,\int_0^1 \D\tau\,\left[Q(-\tau) + Q(\tau) \right]\, Q(0) - \frac12\,\int_0^1 \D\tau \,\left[Q(-\tau)\,Q(1-\tau) + 
	Q(\tau)\,Q(\tau-1)\right] \nonumber  \\[1.5ex]  \label{PUNL5}
	 & = & - \int_0^1 \D\tau \,Q(\tau)\,Q(\tau-1) \,,
\end{eqnarray}
where the vanishing of the mean value of $Q$ has been included. 

Since $Q(\tau)$ corresponds to the positive values of the variable whereas $Q(\tau-1)$ corresponds to the negative ones, both factors in the integral are independent and the integral has no definite sign. 

This can also be seen by direct computation using the Fourier series:
\begin{equation}  \label{PUNL2a}
Q(\tau) = \sum_{n\neq 0} a_n\,e^{i \pi n\tau} \,,\qquad {\rm with} \qquad a_n = \frac12\,\int_\RR \D\tau\, Q(\tau)\, e^{-i\pi n\tau} \,,   
\end{equation} 
where $\,a_{-n} = a_n^\ast \,$ because $q(t)$ is real (the diacritic $^\ast$ means ``complex conjugate''). Substituting the series into equation (\ref{PUNL5}), this becomes
\begin{eqnarray*}
  h &=& - \sum_{n, m \neq 0} a_n\,a_m\,\int_0^1 \D\tau\,e^{i \pi [\tau(m+n)- m]} \\[1.5ex]
	 & = & - \sum_{n\neq -m\neq 0} a_n\,a_m\,\frac{(-1)^n - (-1)^m}{i\,\pi(m+n)} - \sum_{n\neq 0} a_n \,a_{-n}\,(-1)^n  \,.
\end{eqnarray*}	
The first sum vanishes for symmetry reasons, therefore $\,\displaystyle{h = \sum_{n=1}^\infty (-1)^n |a_n|^2  } \,$ and the energy has no definite sign. 
This counterexample proves that the converse of Lagrange Theorem is false.

\section{Conclusion  \label{SC}}

We have revisited the long‑standing claim that higher‑derivative and genuinely nonlocal Lagrangian systems are unavoidably plagued by the so‑called "Ostrogradsky instability". Our analysis, based on a systematic Liapunov approach rather than on energetic arguments alone, demonstrates that this claim is not generally valid. The misconception can be traced to a logical misstep: the fallacy of the inverse of the Lagrange-Dirichlet theorem and treat a sufficient condition for stability as if it were necessary.

We have presented several examples of higher-order and nonlocal Lagrangian systems that are stable, despite the fact that the Hamiltonian is not positive definite. We have focused mainly on variants of the second-order Pais-Uhlenbeck oscillator: free and with interaction, both internal and external, and found that the free case is always stable while the cases with interaction are stable if the coupling parameters belong in a certain range. The case of (finite) higher order-oscillators can be treated in the same way and therefore we have omitted them. We have also studied a genuinely nonlocal system, a PU oscillator of infinite order without interaction, which turns out to be stable. 

Our treatment is based on the Liapunov theory and mainly on Corollary \ref{C1}, which is an extension of the Lagrange-Dirichlet theorem in which the energy is replaced with an integral of motion conveniently chosen that has a strict minimum at the equilibrium point. This is novel from other approaches typically found in the literature. 

\section*{Acknowledgement}
J. Ll. was supported by the Spanish MINCIU and ERDF (project ref. PID2021-123879OB-C22).
The authors also thank C Deffayet and A Vikman for valuable and clarifying discussions concerning their work \cite{Defayet}.

\section*{Appendix A}
Let us consider the linear system 
\begin{equation} \label{A1}
\dot{\mathbf{x}} = \mathbb{M}\,\mathbf{x} \,, \qquad {\rm with} \qquad 
\mathbf{x} = \left(\begin{array}{c}
 x_1 \\ x_2 \\ \vdots 
\end{array} \right)  
\end{equation}
and $\,\mathbb{M}\,$ is a constant square $n\times n$ matrix. This system has one equilibrium point, namely $\,x_k = 0\,,\,\, k = 1\ldots n\,$.

We aim to prove that, provided that the characteristic roots of $\mathbb{M}$ either have negative real part or are imaginary and simple, then the equilibrium point is stable. 
We start introducing the linear transformation
\begin{equation} \label{A2}
 \mathbf{x} = \mathbb{A}\, \mathbf{Y}\,, \qquad  \mathbf{Y} = \mathbb{A}^{-1}\, \mathbf{x} \,, 
\end{equation}
where $\mathbb{A}$ is a constant regular square $n\times n$ matrix. Then, the system (\ref{A1}) becomes 
\begin{equation} \label{A3}
\dot{\mathbf{Y}} = \mathbb{M}^\prime\,\mathbf{Y} \,, \qquad {\rm with} \qquad 
\mathbb{M}^\prime = \mathbb{A}^{-1}\,\mathbb{M}\,\mathbb{A}\,. 
\end{equation} 

Since both transformations are linear and finite dimensional, they are continuous, whence it follows that (\ref{A1}) is stable if, and only if, (\ref{A3}) is stable too. Indeed, introducing the norm
$\, \| \mathbf{x}\|_\infty := \sup \{|x_1 |, \ldots |x_n|\}\,$, we have that
$$ \| \mathbf{Y}(t)\|_\infty \leq n\,\| \mathbb{A}^{-1}\|_\infty \,\| \mathbf{x}(t)\|_\infty
\qquad {\rm and} \qquad 
\| \mathbf{x}(t)\|_\infty \leq n\,\| \mathbb{A}\|_\infty \,\| \mathbf{Y}(t)\|_\infty  \,.$$
The proof follows easily after combining the latter with Definition \ref{D1}, namely
$$ \forall \,\varepsilon > 0 \,, \quad \exists \,\delta \quad\mbox{such that} \quad \| \mathbf{x}(0)\|_\infty < \delta \qquad \Rightarrow\qquad  \| \mathbf{x}(t)\|_\infty < \varepsilon \,, \quad \forall \,t \,.$$

Let us choose $\mathbb{A}$ so that $\mathbb{M}^\prime$ is the canonical Jordan matrix, which is made of as many diagonal blocks as characteristic roots, that is
\begin{equation} \label{A5}
 \mathbb{M}^\prime = \left(\begin{array}{c|c|l}
                              \mathbb{M}_1 & 0 & \ldots \\
															\hline
															 0 & \mathbb{M}_2 & \ldots \\
															\hline
															\multicolumn{2}{c|}{\ldots} & \ddots 
															\end{array}   \right) 
\qquad {\rm and} \qquad 
\mathbf{Y} = \left(\begin{array}{c}
               \mathbf{Y}_ 1 \\
							\hline
							\mathbf{Y}_2  \\
							\hline 
							\vdots
							\end{array}  \right)   \,.
\end{equation}
$\mathbb{M}_\alpha$  is a square $n_\alpha \times n_\alpha$ matrix, where $n_\alpha$ is the dimension of the corresponding invariant subspace, 
$$ \mathbb{M}_\alpha = \left(\begin{array}{cccc}
                              \lambda_\alpha &   & \ldots & \\
															 \sigma_1 & \lambda_\alpha  &  & \vdots \\
															0 & \sigma_2 & \lambda_\alpha &   \\
															 & \ldots & & \ddots
															\end{array}   \right)  \,, \qquad \mbox{where $\sigma_j$ is 0 or 1.}$$
Equation (\ref{A3}) then amounts to 
$$ \dot{\mathbf{Y}}_\alpha = \mathbb{M}^\prime_\alpha\,\mathbf{Y}_\alpha   \,, \quad \forall \,\alpha\,,$$
whose general solution is 
\begin{equation} \label{A4}
\mathbf{Y}_\alpha(t) = e^{t \lambda_\alpha}\,\mathbf{P}(t) \,, \qquad \mbox{where} \qquad \mathbf{P}(t) = \mathbf{P}_0+ t\,\mathbf{P}_1 + t^2\,\mathbf{P}_2 \ldots 
\end{equation}
is a polynomial of degree $n_\alpha -1\,$ and its coefficients, $\,\mathbf{P}_j\,$, depend linearly on $\,\mathbf{Y}_\alpha(0)\,$ up to $n_\alpha -1\,$. 
In case that $\realp(\lambda_\alpha)<0\,$, we have that $\,\|\mathbf{Y}_\alpha(t)\|\,$ goes to zero when $ t\rightarrow\infty\,$.

In case that $\lambda_\alpha = i\,\omega_\alpha\,$ is imaginary and simple, then (\ref{A4}) is simply
$$ \mathbf{Y}_\alpha(t) = e^{i\,t \omega_\alpha}\,\mathbf{Y}_\alpha(0)\qquad {\rm and} \qquad 
\|\mathbf{Y}_\alpha(t)\|_\infty =\|\mathbf{Y}_\alpha(0)\|\,.$$

In any case the definition (\ref{A5}) implies that 
$$ \|\mathbf{Y}(t)\|_\infty = \sup \{ \|\mathbf{Y}_\alpha(t)\| \} $$ 
and therefore, provided that the characteristic roots either have negative real parts or are imaginary and simple, 
\begin{itemize}
\item either $\,\displaystyle{ \lim_{t\rightarrow\infty} \|\mathbf{Y}(t)\|_\infty = 0 } \,$, if no characteristic root is imaginary,
\item or $\,\|\mathbf{Y}(t)\|_\infty \leq \|\mathbf{Y}(0)\|_\infty\,$, in the opposite case.
\end{itemize}
The stability of the equilibrium point then follows straightforward.

\section*{Appendix B}
We shall prove that there are quadratic potentials $\,V= V(q_a) =\displaystyle{\frac12 \,\left(a\, q_1^2+ b \,q_2^2\right) + c \,q_1 q_2 }\,$, with $c\neq 0\,$, such that equation (\ref{PUI2}) admits a solution $\,G(q,p)\,$ that is quadratic and positive. As such, it has a strict minimum at the origin $\,q_a=p_b =0\,$ and it will serve as a Liapunov function to ensure the stability of the equilibrium point.   

Equation (\ref{PUI2}) can be written in the matrix form
\begin{equation}\label{PUI2z}
\mathbb{H}\,\Omega\,\mathbb{G} + \left(\mathbb{H}\,\Omega\,\mathbb{G} \right)^T = 0 \,,
\end{equation}
with
\begin{equation}\label{PUI2a}
 \mathbb{H} =\left(\begin{array}{c|c}
                      \mathbf{A} & 0 \\
											\hline
											0 & \mathbf{B}
											\end{array}\right) \,, \qquad 
	\mathbb{G} =\left(\begin{array}{c|c}
                      \mathbf{U} & \mathbf{R} \\
											\hline
											\mathbf{R}^T & \mathbf{S}
											\end{array}\right)		 \,, 
\qquad 	\Omega =\left(\begin{array}{c|c}
                      0 & \mathbf{I}_2 \\  
											\hline
											- \mathbf{I}_2 & 0
											\end{array}\right)		 \,, 
\end{equation}											
where
\begin{equation}\label{PUI2b}											
\mathbf{A} = \left(\begin{array}{cc}
                     w & c \\
										 c &  v
										\end{array}\right)    \,,\quad \qquad 	
\mathbf{B} = \left(\begin{array}{cc}
                     1 & 0 \\
										 0 &-1
										\end{array}\right)		 \qquad \,, \qquad 	w:=a + \omega_1^2 \qquad \,, \qquad v = b - \omega_2^2  \,,
\end{equation}
 $\,\mathbf{U}\,$ and $\,\mathbf{S}\,$ are symmetric, and the ``boldmath'' type is used to denote square 2$\times$2 matrices.
																			
Substituting them in the matrix equation (\ref{PUI2z}), we obtain:
\begin{equation}\label{PUI3} 
    \mathbf{A} \,\mathbf{R}^T + \mathbf{R} \,\mathbf{A} = 0\,, 		\qquad  
    \mathbf{B} \,\mathbf{R} + \mathbf{R}^T \,\mathbf{B} = 0\,, \qquad 
    \mathbf{A} \,\mathbf{S} - \mathbf{U} \,\mathbf{B} = 0\,. 
\end{equation}
If these equations are to hold for any quadratic potential $V(q)$, from the first two of them it follows that $\,\mathbf{R}=0\,$. As for the third one, it amounts to
\begin{equation}\label{PUI4} 
     \mathbf{U} = \mathbf{A} \, \mathbf{S} \,\mathbf{B} \,, 
\end{equation}
which determines $\,\mathbf{U}\,$ for a given $\, \mathbf{S}\,$. Moreover, as $\,\mathbf{U}\,$ is symmetric, it also implies the following constraint on $\,S\,$
\begin{equation}\label{PUI4a} 
  \mathbf{A} \, \mathbf{S} \,\mathbf{B} -  \mathbf{B} \, \mathbf{S} \,\mathbf{A} = 0 \,. 
\end{equation} 
Since a particular solution of (\ref{PUI2z}) is enough for our purpose, we choose 
$\; S_{11} = S_{22} = 1 \;$ and the condition (\ref{PUI4a}) determines
$$ S_{12} = - \frac{2c^2}{w+v} \,. $$
With this choice,  (\ref{PUI4}) yields
$$ U_{11} = w - \frac{2c^2}{w+v} \,, \qquad
U_{22} = -v + \frac{2c^2}{w+v}  \qquad {\rm and} \qquad U_{12} =\frac{w-v}{w+v}\, c \,. $$
Thus $\,\displaystyle{ G =\frac12\,U_{ab}\, q_a q_b + \frac12\,S_{ab}\, p_a p_b}\,$ is an integral of motion and it has a strict minimum in the equilibrium point $\,p_1=p_2=q_1=q_2=0\,$, which is stable [Corollary \ref{C1}] provided that the Hessian matrix is positive definite, that is
\begin{equation}\label{PUI8} 
\det(\mathbf{S})  > 0 \,,\qquad \quad  w - \frac{2c^2}{w+v}  > 0 	\qquad {\rm and} \qquad  \det(\mathbf{U})  > 0\,.
\end{equation}
The first of these conditions amounts to 
\begin{equation}\label{PUI9} 
 (w+v)^2 > 4 c^2  \,.
\end{equation}
In turn equation (\ref{PUI4}) implies that the third condition is equivalent to $\,\det(\mathbf{A}) < 0\,$, that is
\begin{equation}\label{PUI8a} 
  w\,v  < c^2 \,.
\end{equation} 
Finally, by a thorough analysis, which we will not detail here, and that includes (\ref{PUI9}-\ref{PUI8a}), it can be proved that the second condition (\ref{PUI8}) is equivalent to
\begin{equation}\label{PUI8c} 
 w - v > 0 \,.
\end{equation}

Summarizing, as long as the coefficients $a$, $b$ and $c$ of the quadratic potential $V(q)$ satisfy the inequalities (\ref{PUI9}-\ref{PUI8c}), or
equivalently   
\begin{equation}\label{PUI101} 
   (a + b +\omega_1^2- \omega_2^2)^2 > 4\, c^2 > 4 \,(a + \omega_1^2) \,(b - \omega_2^2) 
	 \qquad {\rm and } \qquad a + \omega_1^2 > b - \omega_2^2  \,,
\end{equation} 
the integral of motion 
\begin{equation}\label{PUI102} 
   G = h_1 + h_2 + V - \frac{c^2 q_1^2 - \left[c^2 + b(a + b +\omega_1^2 -\omega_2^2)\right]\,q_2^2 + 2 c \left[b-\omega_2^2 \right] q_1 q_2 + 2 c p_1 p_2 }{a + b +\omega_1^2- \omega_2^2} \,
\end{equation} 
warrants, by Corollary \ref{C1}, that the equilibrium point $\,p_1=p_2=q_1=q_2=0\,$ is stable.

\input{Appendix_C}

\printbibliography

\end{document}

%% file: GhostsNoRunaways.tex
\subsection{A double oscillator with ghosts and without runaways or instabilities}
Deffayet et al \cite{Defayet,Defayet2} have studied a double oscillator system of the sort (\ref{FPU5}), with $\omega_1=\omega_2\,$ and a conveniently chosen interaction potential $V$. They prove that, although the energy is not bounded from below, the system is stable in the sense of Liapunov (Definition \ref{D1}).

Consider the double-oscillator Hamiltonian (\ref{PUI0}), with $\omega_1 = \omega_2\,$ that we take equal to 1 by a suitable scaling of time, perturbed by an internal interaction term: 
\begin{equation}\label{PUI10}     
 H = h_1-h_2 + V(q) \,, \qquad h_a =\frac12\,\left(p_a^2 + q_a^2 \right) \,.
\end{equation}
Notice that the system cannot be derived from a perturbed PU oscillator because the latter leads to a Hamiltonian like this only if $\omega_1 \neq \omega_2$; otherwise the transformations (\ref{FPU4}-\ref{FPU4a}) do not apply.
The reason why we study these systems here, despite of the fact that they cannot be derived from a higher-order Lagrangian, is to demonstrate that the quest of a Liapunov function is a powerful tool to find other interaction potentials that leave the system stable, thus providing a systematic way of selecting more general functions $V$. 

The case $\omega_1 = \omega_2=1$ is manageable because the free system has a third algebraic integral of motion, namely $\,k = p_1 q_2 + p_2 q_1 \,$, which facilitates the search of integrals of motion of the perturbed system. We try with  
\begin{equation}\label{PUI10z}     
 G = h_1+h_2 + k^2 +W(q) \,, 
\end{equation}
where $W(q)$ is some function to be determined. The condition $\,\{H, G\} = 0\,$ easily leads to
$$ p_1\,\left(- \partial_1 W + \partial_1 V + 2\,q_2 \,\left[q_2 \partial_1 V + q_1 \partial_2 V\right]  \right) +
 p_2\,\left( \partial_2 W + \partial_2 V + 2\,q_1 \,\left[q_2 \partial_1 V + q_1 \partial_2 V\right]  \right) = 0 \,,  $$
where $\displaystyle{\partial_b W:= \frac{\partial W}{\partial q^b}\,}$ and so on. 
Since it must hold for all values of $p_b$, it implies that
\begin{equation}\label{PUI11}     
\partial_1 W = \left(1 + 2 q_2^2 \right)\, \partial_1 V + 2 q_1 q_ 2 \partial_2 V   \qquad {\rm and} \qquad 
- \partial_2 W = \left(1 + 2 q_1^2 \right)\, \partial_2 V + 2 q_1 q_ 2 \partial_1 V \,.
\end{equation}

Introducing the new variables
$$ u = 1 + 2\,\left(q_1^2+q_2^2\right) \,, \qquad v =  2\,\left(q_1^2 - q_2^2\right) \,,$$
the partial derivatives transform as
$\, \partial_1 =4 q_1\,\left(\partial_u + \partial_v\right) \,$  and $\, \partial_2 =4 q_2\,\left(\partial_u - \partial_v\right) \,$,
and equations (\ref{PUI11}) become
\begin{equation}   \label{PUI12}     
\partial_v W = u\,\partial_u V \qquad {\rm and} \qquad  \partial_u W = \partial_v V - v\,\partial_u V \,.
\end{equation}
The integrability condition then yields a PDE on $V(u,v)\,$:
\begin{equation}   \label{PUI12z}     
u\,\partial_{uu}V + v\,\partial_{uv}- \partial_{vv}V + 2 \partial_u V = 0 \,,
\end{equation}
that is, for any potential $V$ fulfilling this equation, a solution $W$ of (\ref{PUI12}) exists, and the function $G$ defined by (\ref{PUI10z}) is an integral of motion of the system and is a candidate for Liapunov function.  

In Appendix C we solve equations (\ref{PUI12z}) and  (\ref{PUI12}) to prove that the general potential admitting an integral of motion like (\ref{PUI10z}) is 
\begin{equation}   \label{PUI13A}     
V(u,v) = \frac1{\Delta}\,\left[\Phi\left(v + \Delta \right) + \Phi\left(v -\Delta \right)\right]
\end{equation}
with
\begin{equation}
 v =  2\,\left(q_1^2 - q_2^2\right) \quad {\rm and} \quad \Delta:= 2\,\sqrt{\left(q_1^2-q_2^2\right)^2 + 2 \left(q_1^2+q_2^2\right) + 1}\,,   
\end{equation} 
where $\Phi $ is an arbitrary function. We also prove there that (\ref{PUI10z}) is an integral of motion with
\begin{equation}   \label{PUI13B}     
W = - \frac1{2 \Delta}\,\left[\left(v - \Delta \right)\, \Phi\left(v + \Delta \right) + \left(v + \Delta \right)\,\Phi\left(v -\Delta \right)\right]\,. 
\end{equation}
Notice that the potential proposed in ref. \cite{Defayet} corresponds to the choice $\Phi = {\rm constant}\,$. 

Now the equations for the Hamiltonian (\ref{PUI10}) are 
$$ \dot q_b = \sigma_b p_b \,, \qquad \dot p_a = - \sigma_a q_a - \partial_a V \,, \qquad \sigma_1 = -\sigma_2 = 1 \,,$$
and the equilibrium points are the solutions of $\; p_b = 0 \;$ and $\; q_a\,\left(1 + 4 \left[\partial_v V + \sigma_a \partial_u V  \right]  \right) = 0 \,$, that is $\; p_1 = p_2 = 0 \;$ and
\begin{enumerate}
\item $\;\;q_1=q_2=0\,$,
\item $\;\;q_1=0\,$ and $ \; 1 + 4 \,\left(\partial_v V - \partial_u V  \right) = 0\,$,
\item $\;\;q_2=0\,$ and $ \; 1 + 4 \,\left(\partial_v V + \partial_u V  \right) = 0\,$ and 
\item $\;\;1 + 4 \,\partial_v V = \partial_u V = 0\,$.
\end{enumerate}

For an equilibrium point to be stable, it is sufficient that the integral of motion $G$ has a strict minimum, that is the partial derivatives of $G(p,q)$ must vanish and the Hessian matrix must be positive definite. We easily obtain that
$$ \frac{\partial G}{\partial p_a} = p_a + 2 k q_{a^\prime} \,, \qquad a^\prime \neq a \,,\qquad {\rm and}  \qquad
 \frac{\partial G}{\partial q_b} = q_b \,\left(1 + 4 \left[\partial_v W + \sigma_a \partial_u W  \right]  \right) \,.$$

We will focus on the stability of the equilibrium point $\,p_a=q_b=0\,$, which is obviously a stationary point of $G$. Moreover, the only non-vanishing elements of the Hessian matrix are:
$$ \frac{\partial^2 G}{\partial p^2_a} = 1 \qquad {\rm and}  \qquad \frac{\partial^2 G}{\partial q_a \partial q_b} = \delta_{ab} \,\left(
1 + 4 \left[\partial_u W + \sigma_b \partial_v W \right] \right)  \,, $$ 
and therefore the equilibrium point is stable if $\; 1 + 4 \partial_u W \pm 4 \partial_v W > 0 \,$.
Including (\ref{PUI13B}) and the fact that in this equilibrium point $ v=0 $ and $ \Delta = 2\,$, this turns out to be equivalent to
$$ 1 + 4 \Phi^\prime(2) > \Phi(2)+\Phi(-2) >  4 \Phi^\prime(-2) - 1 \,. $$
Summarizing, for an arbitrary function $\Phi$ fulfilling the above inequality, the potential (\ref{PUI13A}) has a stable equilibrium point in $\,p_a=q_b=0\,$, despite the energy is not bounded from below.

%% file: Appendix_C.tex
\section*{Appendix C: The Liapunov function (\ref{PUI10z}) }
Consider the system (\ref{PUI11})
$$ \partial_1 W = \left(1 + 2 q_2^2 \right)\, \partial_1 V + 2 q_1 q_ 2 \partial_2 V   \qquad {\rm and} \qquad 
- \partial_2 W = \left(1 + 2 q_1^2 \right)\, \partial_2 V + 2 q_1 q_ 2 \partial_1 V \, $$
and introduce the new variables $\, u = 1 + 2\,\left(q_1^2+q_2^2\right) \,, \quad v =  2\,\left(q_1^2 - q_2^2\right) \,$. 
The partial derivatives transform as
$\, \partial_1 =4 q_1\,\left(\partial_u + \partial_v\right) \,$  and $\, \partial_2 =4 q_2\,\left(\partial_u - \partial_v\right) \,$,
equations (\ref{PUI11}) become
\begin{equation}   \label{PUI12m}     
\partial_v W = u\,\partial_u V \qquad {\rm and} \qquad  \partial_u W = \partial_v V - v\,\partial_u V \,.
\end{equation}
and the integrability condition then yields a PDE on the interaction potential:
\begin{equation}   \label{PUI12n}     
u\,\partial_{uu} V+ v\,\partial_{uv} V- \partial_{vv} V+ 2 \partial_u V= 0 \,,
\end{equation}
that is, for any potential $V$ fulfilling this equation, a solution $W$ of (\ref{PUI12m}) exists and $G$ defined by (\ref{PUI10z}) is an integral of motion of the system and is a candidate for Liapunov function.  

We now solve this integrability equation by Fourier transform methods and define 
$$ \tilde{V}(\xi,\zeta) = \frac1{2\pi}\,\int_{\RR^2} \D u\,\D v\,V(u,v)\,e^{i(\xi u+\zeta v) } \,. $$
The FT of equation (\ref{PUI12n}) is 
$$ \left(\xi \partial_\xi + \zeta \partial_\zeta \right) \tilde{V} + \left(1 - \frac{i \,\zeta^2}\xi \right)\,\tilde{V} = 0 \,,$$
whose general solution is 
\begin{equation}   \label{PUI13}     
\tilde{V}\left(\frac\zeta\xi , \xi\right) = \varphi\left(\frac\zeta\xi\right)  \,\frac1\xi\, e^{i \zeta^2/\xi} \,,
\end{equation}
where $\varphi$ is an arbitrary function. Then, performing the inverse Fourier transform we obtain that
$$ V(u,v) = \frac1{\rho_+-\rho_-}\,\left[\varphi(\rho_+) + \varphi(\rho_-)\right]\,, \qquad 
\rho_\pm = \frac12\,\left(v \pm \sqrt{v^2 + 4 u} \right) \,  $$
and, taking $\,\Phi(s) := \varphi(s/2)\,$, we finally have:
\begin{equation}   \label{PUI13a}     
V(u,v) = \frac1{\sqrt{v^2 + 4 u}}\,\left[\Phi\left(v + \sqrt{v^2 + 4 u} \right) + \Phi\left(v - \sqrt{v^2 + 4 u} \right)\right]\,. 
\end{equation}
$\Phi$ is an arbitrary function and taking $\Phi = {\rm constant}$ yields the potential in ref. \cite{Defayet}.

In terms of the variables $v$ and $\Delta=\sqrt{v^2 + 4 u}\,$, the potential is
\begin{equation}   \label{PUI13b}     
V(\Delta,v) = \frac1{\Delta}\,\left[\Phi\left(v + \Delta \right) + \Phi\left(v - \Delta \right)\right]\,. 
\end{equation}

In terms of these variables the first equation in the differential system (\ref{PUI12m}) becomes
$$ \partial_v\left(W +\frac{v}2\,V\right) = \partial_\Delta \left(\frac{\Delta}2\,V\right) = \frac{\Phi^\prime(v+\Delta)-\Phi^\prime(v+\Delta)}2 = \partial_v\left(\frac{\Phi(v+\Delta)-\Phi(v+\Delta)}2 \right) \,,$$
whence it follows that
\begin{equation}   \label{PUI13c}     
W(u,v) = -\frac1{2 \Delta}\,\left[\left(v - \Delta \right)\, \Phi\left(v + \Delta \right) + \left(v + \Delta \right)\,\Phi\left(v - \Delta \right)\right]\,. 
\end{equation}